%% file: SPLC2016.tex
\documentclass{sig-alternate-05-2015}
\usepackage{graphicx}
\usepackage{xcolor}
\usepackage{cite}
\usepackage[font=small,bf]{caption}

\usepackage{color}
\usepackage{xcolor}
\usepackage{float}

\usepackage{upgreek}
\usepackage{fixltx2e}
\usepackage{listings}

\newcommand{\mymark}[1]{\textsuperscript#1}

\newcommand{\LinuxConfigOptions}{14,000}
\newcommand{\LinuxConfigOptionsOurAnalysis}{8,537}
\newcommand{\NRreportedCVE}{1,314} 
\newcommand{\NRFilesLinuxKernel}{12,075}
\newcommand{\NRParsedFilesLinuxkernel}{11,956}

\everypar{\looseness=-1.5}
\makeatletter
\newcommand*{\NoBreakPar}{\vspace{\baselineskip}\par\nobreak\@afterheading}
\makeatother

\makeatletter
\def\@copyrightspace{\relax}
\makeatother

\begin{document}
\input{SPLC2016_complete_generated.tex}

\title{Do \#ifdefs Influence the Occurrence of Vulnerabilities? \\An Empirical Study of the Linux Kernel}	

\numberofauthors{1} 
\author{
\alignauthor Gabriel Ferreira\mymark{1}, Momin Malik\mymark{1}, Christian K\"{a}stner\mymark{1}, J\"{u}rgen Pfeffer\mymark{1}, Sven 
Apel\mymark{2}\\
\mymark{1}\affaddr{Carnegie Mellon University, United States}\\
\mymark{2}\affaddr{University of Passau, Germany}\\
}

\maketitle
\begin{abstract}
Preprocessors support the diversification of software products with \#ifdefs, but also require additional effort from developers to 
maintain and understand variable code. We conjecture that \#ifdefs cause developers to produce more vulnerable code because they are 
required to reason about multiple features simultaneously and maintain complex mental models of dependencies of configurable code.

We extracted a variational call graph across all configurations of the Linux kernel, and used configuration complexity metrics to compare 
vulnerable and non-vulnerable functions considering their vulnerability history. Our goal was to learn about whether we can observe a 
measurable influence of configuration complexity on the occurrence of vulnerabilities. 

Our results suggest, among others, that vulnerable 
functions have higher variability than non-vulnerable ones and are also constrained by fewer configuration options. This suggests that 
developers are inclined to notice functions appear in frequently-compiled product variants. We aim to raise developers' awareness to 
address variability more systematically, since configuration complexity is an important, but often ignored aspect of software product 
lines. 
\end{abstract}

%
%


%
%

%
%



\setcounter{section}{0}
\section{Introduction}
Diversification of software products is widely desired, but also induces challenges in development and maintenance processes of software 
product lines \cite{pohl:2005:SPL, liebig:2010}. Preprocessor directives (\#ifdef statements) are frequently used as a mechanism to support 
code variability and thereby permit the diversification of software products. However, it is known that the presence of \#ifdefs in source 
code complicates maintenance tasks and requires additional effort from developers when trying to understand feature code dependencies 
\cite{liebig:2010, erwig:2011, schulze:2013}.
  
In this paper, we define \emph{configuration complexity} as the complexity induced by the presence of \#ifdefs in the code, and we 
conjecture that it causes developers to make mistakes that lead to more vulnerable code. Our assumption is motivated by the observation 
that humans have a limited capacity to keep an accurate and complete mental model of code dependencies \cite{latoza:2006}. When 
considering the scenario of maintaining multiple software products and reasoning about many variants simultaneously, this limitation could 
result in serious consequences. For example, it could cause unexpected feature interactions and feature code to be inadvertently executed 
or bypassed, creating opportunities for attackers to exploit software systems \cite{nhlabatsi:2008}. 

\begin{figure}[t]
	\centering
	\includegraphics[width=.8\linewidth, height=8cm, clip=true]{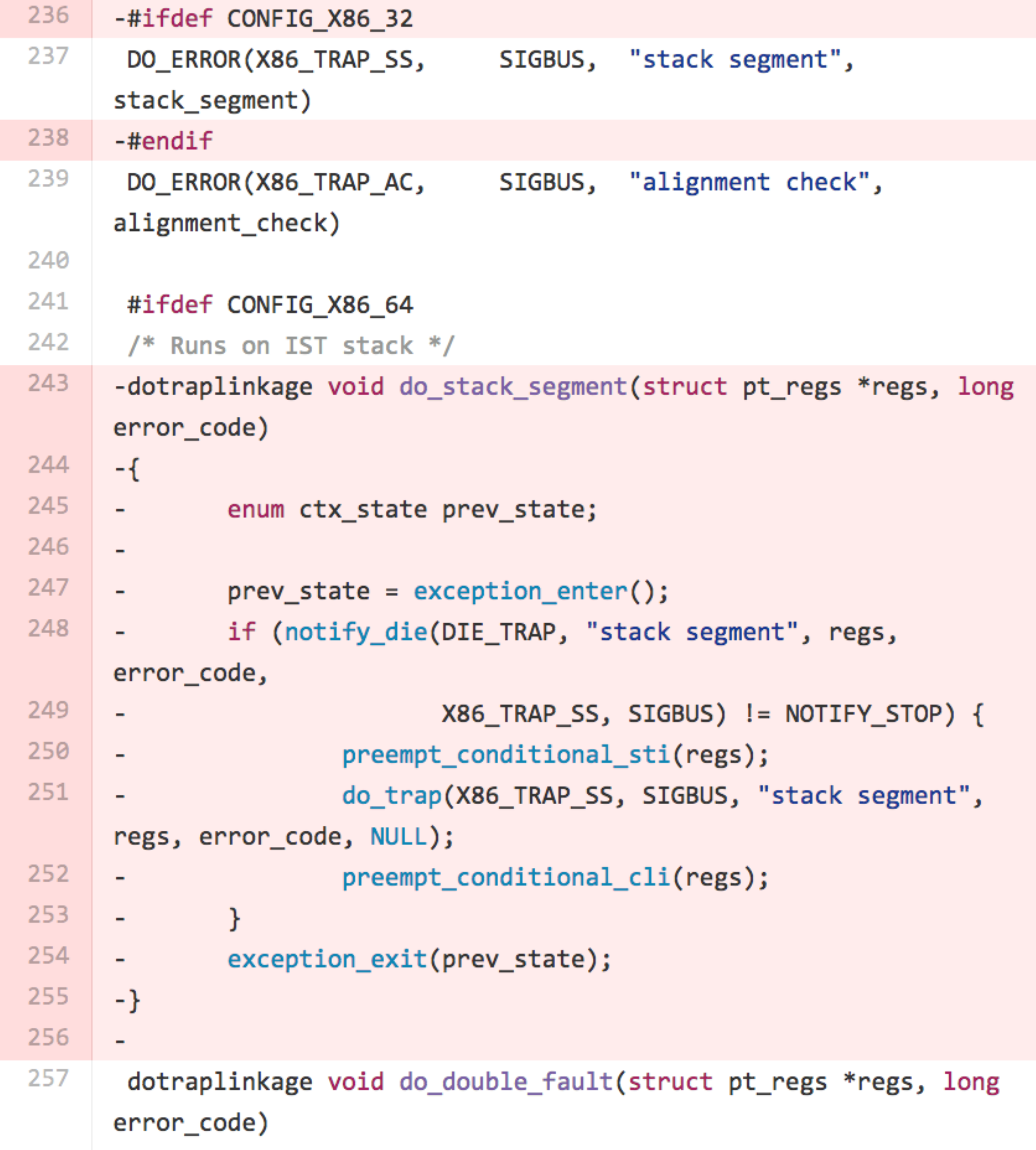}
	\caption{\small\bfseries{Snippet from the commit diff that fixed the \mbox{CVE-2014-9322} vulnerability 
	\textit{(arch/x86/kernel/traps.c)}}}
	\label{cve_example}
\end{figure}

Figure \ref{cve_example} shows a snippet of a commit diff that fixed a vulnerability in file \textit{arch/x86/kernel/traps.c} of the 
Linux kernel, a large configurable software system that shares many characteristics with industrial software product lines 
\cite{sincero:2007, hunsen:2015}. In this example, the \#ifdef statement is used to constrain feature code according to the setting of two 
configuration options. Those are usually Boolean variables that represent features available in a product line and can be enabled or 
disabled in the application engineering process. In this example, function \textit{do\_stack\_segment} is constrained by option 
\textit{CONFIG\_X86\_64}, meaning that it will be compiled and be part of a product variant only when \textit{CONFIG\_X86\_64} 
is enabled.

We seek to characterize configuration complexity of functions and analyze whether it associates with their past vulnerable 
behavior. To this end, we extract and quantify presence conditions (predicates over configuration options) from functions of the Linux 
kernel and use it as a baseline to compare samples of vulnerable and non-vulnerable functions. More than defining and quantifying 
configuration complexity, we aim at understanding whether the configuration aspect of unpreprocessed source code (i.e., the presence of 
\#ifdefs in source code) provides additional information when used in combination with traditional size and structural complexity metrics 
\cite{mccabe:1976, fenton:1998, newman:2010}. 

Ultimately, we are interested in learning whether vulnerable and non-vulnerable functions have  each distinguishable complexity 
characteristics that would potentially allow us to warn developers about critical pieces of a product line. We pose the following research 
questions:
\NoBreakPar
\begin{itemize}
	\item[\textbf{RQ1}]
	\emph{Does configuration complexity associate with past vulnerable behavior of functions?} 
	\item[\textbf{RQ2}]
	\emph{Does configuration complexity provide additional insights about past vulnerable behavior of functions when compared 
		to size and structural complexity?} 
\end{itemize}

Our general hypothesis is that complexity metrics can help maintainers to identify vulnerability-prone code in configurable code. High 
configuration complexity can be used as a warning sign and, in concert with other quality indicators, could help to identify 
potential vulnerabilities, an important facet of what makes software difficult to assure. 

Compile-time configuration complexity has not been considered in analyses before, because existing tools work on preprocessed 
code, that is, in a single configuration after running preprocessor and compiler. Even parsing unpreprocessed code soundly was a challenge 
that was only recently solved \cite{gazzillo:2012, kastner:2011}. Our infrastructure allows, for the first time, to parse (and type check) 
unpreprocessed code, while generating the call graph for all configurations of the Linux kernel. The produced \emph{variational} call graph 
is an important basis for our analysis of configuration complexity of functions.

To define configuration complexity, we design three simple configuration complexity metrics (Section \ref{sec:configuration_comp_analysis}) 
that capture the complexity induced by the presence of \#ifdefs in the code and three structural metrics that capture information on the 
relationship of functions in a call graph (Section \ref{sec:structural_comp_analysis}).

Our results show that vulnerable and non-vulnerable functions have distinct characteristics regarding configuration complexity that can add 
additional value to traditional size and structural-complexity measures \cite{fenton:1998}. For instance, we found that vulnerable 
functions have, on average, three times more \#ifdef statements inside a function than non-vulnerable functions, an effect size greater 
than observable from studying size metrics only. For other measures of configuration complexity, we found similarly encouraging results. 
Our results provide a basis towards the development of prediction models, but more importantly, raise awareness of product-line developers 
to address variability more systematically (for example, with testing \cite{garvin:2011, pohl:2006:SPLTesting} and variability-aware 
analysis \cite{abal:2014, thum:2014}).

Overall, we make the following contributions: 
\begin{enumerate}
	\item We define configuration complexity and provide an infrastructure to measure it on unpreprocessed C code.
	\item We analyze how configuration complexity is associated with past vulnerable behavior of functions and investigate potential 
	confounding effects between our metrics and traditional complexity metrics.
	\item We discuss the general implications of our results for developers and maintainers of product lines.
\end{enumerate}

\section{Background and Motivation}
\label{sec:motivation}
Based on existing work that suggests that the presence of \#ifdefs in source code complicates maintenance tasks and requires additional 
effort from developers when trying to understand variable code dependencies \cite{schulze:2013, erwig:2011, lohmann:2006}, we aim at 
investigating the influence that code configurability has on the occurrence of vulnerabilities. 

Configuration-related vulnerabilities arise in practice and, generally, can have serious consequences. One famous example is 
the Heartbleed vulnerability in OpenSSL (CVE-2014-0160), which affected servers, browsers, and many other systems that use this encryption 
library to secure Internet communication. In this specific case, the vulnerability was associated to one enabled-by-default configuration 
option that was frequently included in the build process, but rarely needed by users of the library.

Another example of a configuration-induced vulnerability was reported for the Linux kernel (CVE-2014-9322). In this case, the code 
responsible for handling stack segment violations was distinct for different computer architectures (32-bit and 64-bit) and caused the 
64-bit version to be vulnerable. Part of the solution to fix this vulnerability involved modifying the file \emph{arch/x86/kernel/traps.c} 
by removing both the specialized function \textit{do\_stack\_segment} responsible for handling error in 64-bit architectures (Lines 243 to 
256) and the \#ifdef directives responsible for applying the default error handling only to 32-bit architectures (Lines 236 and 238).

It has been observed that product line maintainers usually maximize the functionality of systems to reduce the high engineering costs 
required to certify every possible product variant that can be generated \cite{tartler:2012} and also rely on default values for 
configuration options to avoid the burden of reason about the complexity induced by code configurability \cite{georgiev:2012}. 
The latter is even more dangerous because it increases the attack surface of software systems and, potentially, the number of undesired 
interactions among features \cite{nhlabatsi:2008}.

To analyze \#ifdefs, our analysis focuses on compile-time variability to enable systematic reasoning of code configurability 
\cite{post:2008, vonRhein:2016}, rather than relying on sometimes useful, but unsound approximations, such as maximizing the configuration 
options enabled for a product or translating \#ifdefs to if statements \cite{vonRhein:2016}. Moreover, it allows us to explore knowledge 
about configuration options that is sometimes buried in build files and macros, which makes it harder for developers to reason about its 
true effects without performing an in-depth analysis of unpreprocessed code.

The goal of our study was to use simple metrics that could capture our intuition of configuration complexity and allow us to search for 
evidence that the complexity induced by \#ifdefs and configuration options associates with vulnerable behavior of functions. We define and 
operationalize the metrics in more detail in Sections \ref{sec:configuration_comp_analysis} and \ref{sec:structural_comp_analysis}.

One simple metric that we used to capture configuration complexity is the number of internal \#ifdefs that appear inside a function. 
Intuitively, this metric translates to how many times a maintainer would need to switch context between feature blocks (in addition to the 
conditional branches in the code), while trying to understand or modify a piece of configurable code.

Although this and other metrics are simple (Section \ref{sec:configuration_comp_analysis}), we believe they complement traditional 
size and structural complexity metrics, by augmenting individual function properties with other properties that originate from their 
interaction with other functions and their inherent variability.


When analyzing structural complexity, we aim at studying the phenomena that emerge from the interaction of program elements 
\cite{johnson:2009}. To this end, we extract a call graph from unpreprocessed code (Section \ref{sec:variational_cg}) and maintain 
information about its variability by labeling functions (nodes) and function calls (edges). These labels, representing configuration 
complexity, are later quantified and used in combination with other numerical graph-based metrics. Ultimately, we expect to increase the 
usefulness of the traditional metrics \cite{fenton:1998}.


\section{Experimental Setup}
\label{sec:experiment_setup}
We decided on the Linux kernel (version 3.19, x86 architecture) as the subject of our study because it is one of the largest and most 
configurable product line publicly available for analysis \cite{liebig:2010} and, at the same time, one with the most reported 
vulnerabilities. With more than \LinuxConfigOptions{} configuration options available, the Linux kernel is widely used in industry 
and its use expands from high-end servers to mobile phones. From about \LinuxConfigOptions{} configuration options, only 
\LinuxConfigOptionsOurAnalysis{} affect our analysis of the x86 architecture.

In our experiment, we analyze the vulnerability history of functions, by checking whether a certain function has been touched by developers 
to fix past vulnerabilities. Next, we compute configuration complexity metrics for each function and analyze differences in samples of 
vulnerable and non-vulnerable functions along the selected metrics, such as the number of internal \#ifdefs. To avoid fishing for results, 
we carefully design our metrics based on our understanding of how configuration complexity might affect developers; we discuss the metrics 
and their rationale in Sections \ref{sec:configuration_comp_analysis} and~\ref{sec:structural_comp_analysis}.

In addition, we analyzed whether the configuration aspect of the code provides additional information when used in combination with 
traditional size and structural complexity metrics \cite{fenton:1998}. Our analysis considers the potential confounding effects of 
traditional complexity metrics and aims at understanding, quantifying, and isolating the real effect that \#ifdefs have on the complexity 
of variable code. 

\subsection{Mining Vulnerabilities}
\label{sec:mining_vulnerab}
To learn about whether configuration complexity of functions is associated with the occurrence of vulnerabilities, we needed to identify 
which functions have been vulnerable in the past. For this purpose, we mined reported known vulnerabilities from the National Vulnerability 
Database (NVD)\footnote[1]{\small{\url{https://nvd.nist.gov/}}}. This database catalogs information about real vulnerabilities that have 
been reported by developers and users when an exploit had been identified, each given a unique CVE number.
	
When investigating vulnerabilities, we collect information about the commits that have been assigned as responsible for fixing the code 
that was vulnerable in the past. That is, whenever code has been committed to fix a vulnerability, we identify all files and functions that 
have been modified in the commit. 

From the vulnerability database, we identified \NRreportedCVE{} vulnerabilities reported from 1999 to 2015. For each vulnerability 
reported, we collected links to the commits fixing the vulnerability in the source code in either GitHub or kernel.org, resulting in a list 
of commits fixed reported vulnerability. For each commit in the list, we identify all files and functions that have been touched to fix a 
vulnerability. 

We automated the extraction process to reduce human error. Specifically, we download the files from the commit diffs available in the 
commit, parse the C files using \emph{srcML}~\cite{collard:2003}, and then collect information about the location and boundaries of each 
function in the file (begin and end lines). Next, we use the function location to identify whether the changes have been made within the 
limits of a function. The result of this step is a list of functions that have been changed to fix each of the vulnerable files. 
	
To increase our confidence in the extraction process and in the data we were extracting, we also decided to mine the history of 
commit messages directly from the Linux kernel git repository. We observed that some old reported vulnerabilities are not linked to git 
commits, so we assumed we could find information about vulnerable functions directly from commits that are not referenced in the 
vulnerability database. When analyzing the Linux kernel source code repository, we searched through the git history and looked for 
`\emph{CVE-}' identifiers in commit comments. For each commit that matched with our search, we identified files and functions that have 
been modified by analyzing the textual diffs. We manually checked a few instances of the mined CVEs for correctness.
	
In total, we collected information on \NRreportedCVE{} CVEs, successfully parsed \NRParsedFilesLinuxkernel~ files out of 
\NRFilesLinuxKernel~files and extracted \NRFunctionsLinuxKernel~functions of the Linux kernel (x86 architecture). From the set of extracted 
functions, \NRVulnerableFunctionsLinuxKernel~were associated with CVEs and are considered vulnerable; the remaining 
\NRNonVulnerableFunctionsLinuxKernel~functions are considered non-vulnerable. 
	
\subsection{Variational Call Graphs}
\label{sec:variational_cg}

We use variational call graphs to analyze the interactions among functions \cite{johnson:2009} and to investigate how configuration 
complexity influences those interactions. They serve as a technical basis for our structural complexity analysis (Section 
\ref{sec:structural_comp_analysis}).

A call graph is an abstraction of a program that represents potential calls among functions at runtime. Although compact, call graphs are 
relatively cheap to compute and, yet, powerful abstractions of a program's behavior \cite{ferreira:2015}. Besides being beneficial for 
developers to reasoning about software systems, call graphs are a useful approximation of a program's execution, which makes them 
potentially relevant to perform security-related program analysis \cite{wagner:2001}.

\begin{figure}
	\centering
	\includegraphics[width=.9\linewidth]{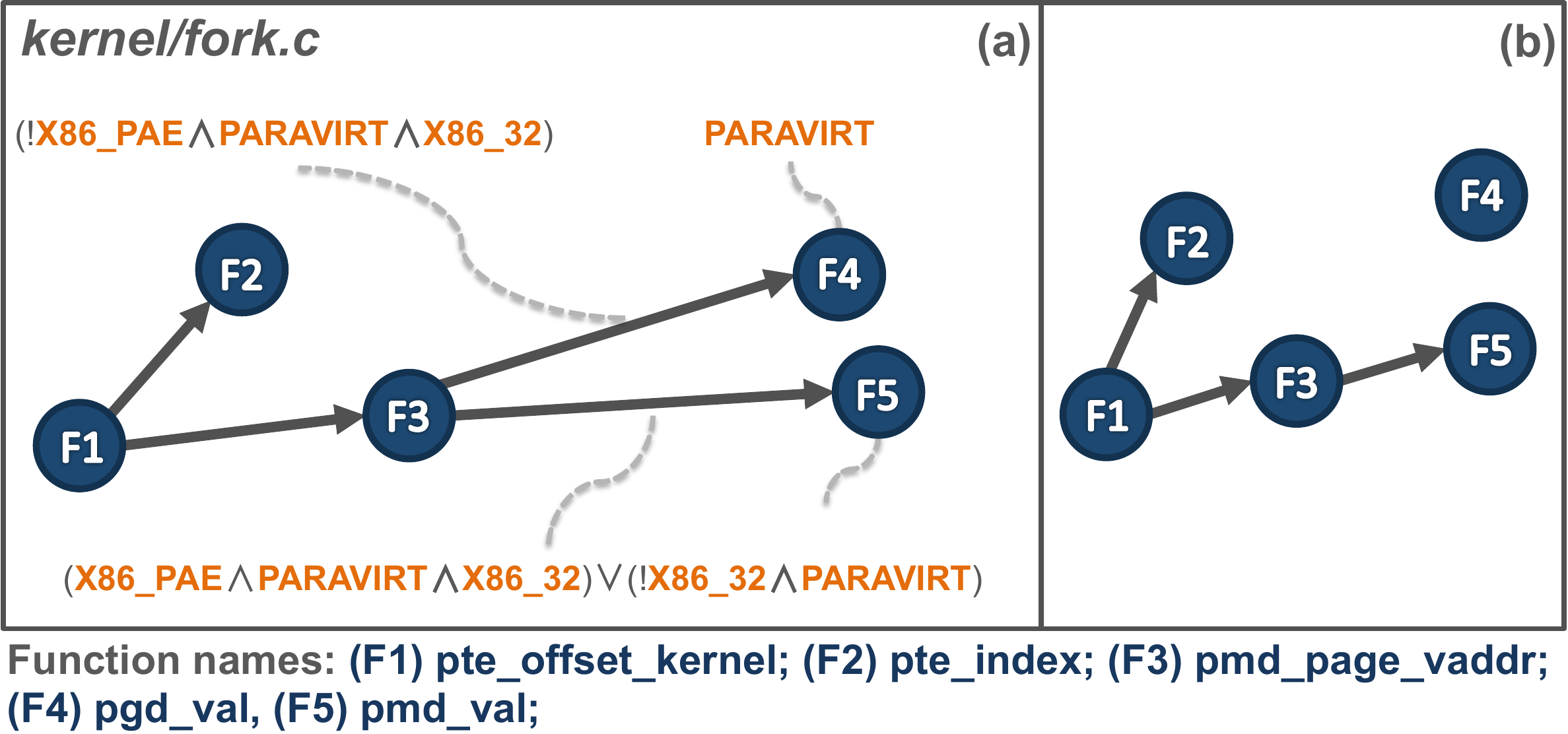}
	\caption{\small\bfseries{Excerpt of a variational call graph extracted from the Linux kernel \textit{kernel/fork.c} file (a) and the 
	resulting call graph when the configuration option \textit{X86\_PAE} is not selected (b). The presence conditions on nodes and edges in 
	(a) show in which condition a function or function call would be included in a product variant.}}
	\label{fig:cg_and_var_cg}
\end{figure}

To take configuration complexity into consideration, we extend the notion of call graph to make a variational call graph that compactly 
represents all possible function definitions and function calls of a given product line. The variational call graph provides the basis to 
analyze the effect of configuration complexity in graph-based metrics. Instead of producing a call graph for each individual system 
configuration, a variational call graph includes all possible nodes and edges of any system configuration, but labels each node and each 
edge with a \emph{presence condition}, characterizing precisely in which configurations a function definition or function call would be 
included \cite{liebig:2013}. The result is a labeled graph that can be used for subsequent analysis; when analyzing configuration 
complexity, we are especially interested in these labels. Figure \ref{fig:cg_and_var_cg}(a) shows an excerpt of the variational call graph 
for the file \textit{kernel/fork.c} and the resulting call graph of a product variant when the configuration option \textit{X86\_PAE} 
is not selected (b).



To analyze configuration complexity, we first computed a variational call graph from the unpreprocessed source code of the Linux 
kernel (version 3.19, x86). To compute it, we implemented our analysis on top of the TypeChef infrastructure \cite{kastner:2011, 
kastner:2012}, which can parse unpreprocessed C code, including preprocessor directives, into a variational Abstract Syntax Tree (AST) 
representing all configurations. The nodes in the AST representation store the configuration information in the 
form of choice nodes \cite{kastner:2011}. By walking over the variational AST, we are able to identify function definitions and function 
calls that occur in the Linux kernel, as well as the presence conditions under which they are enabled or disabled from a product variant. 
To increase the accuracy of the call graph extraction, we implemented a relatively inexpensive but precise pointer analysis 
\cite{ferreira:2015}\footnote{\small{\url{https://github.com/ckaestne/TypeChef/}}}. 
		
\subsection{Null-hypothesis testing}
\label{sec:null_hyp_testing}
The purpose of the tests is to check whether the samples of vulnerable and non-vulnerable functions are different according to the selected 
metrics that we will discuss in Sections \ref{sec:configuration_comp_analysis} and \ref{sec:structural_comp_analysis}. The null hypothesis 
for all tests is that both vulnerable and non-vulnerable functions are drawn from the same distribution of the metric.

\begin{figure*}[!t]
	\centering
	\includegraphics[width=.8\linewidth]{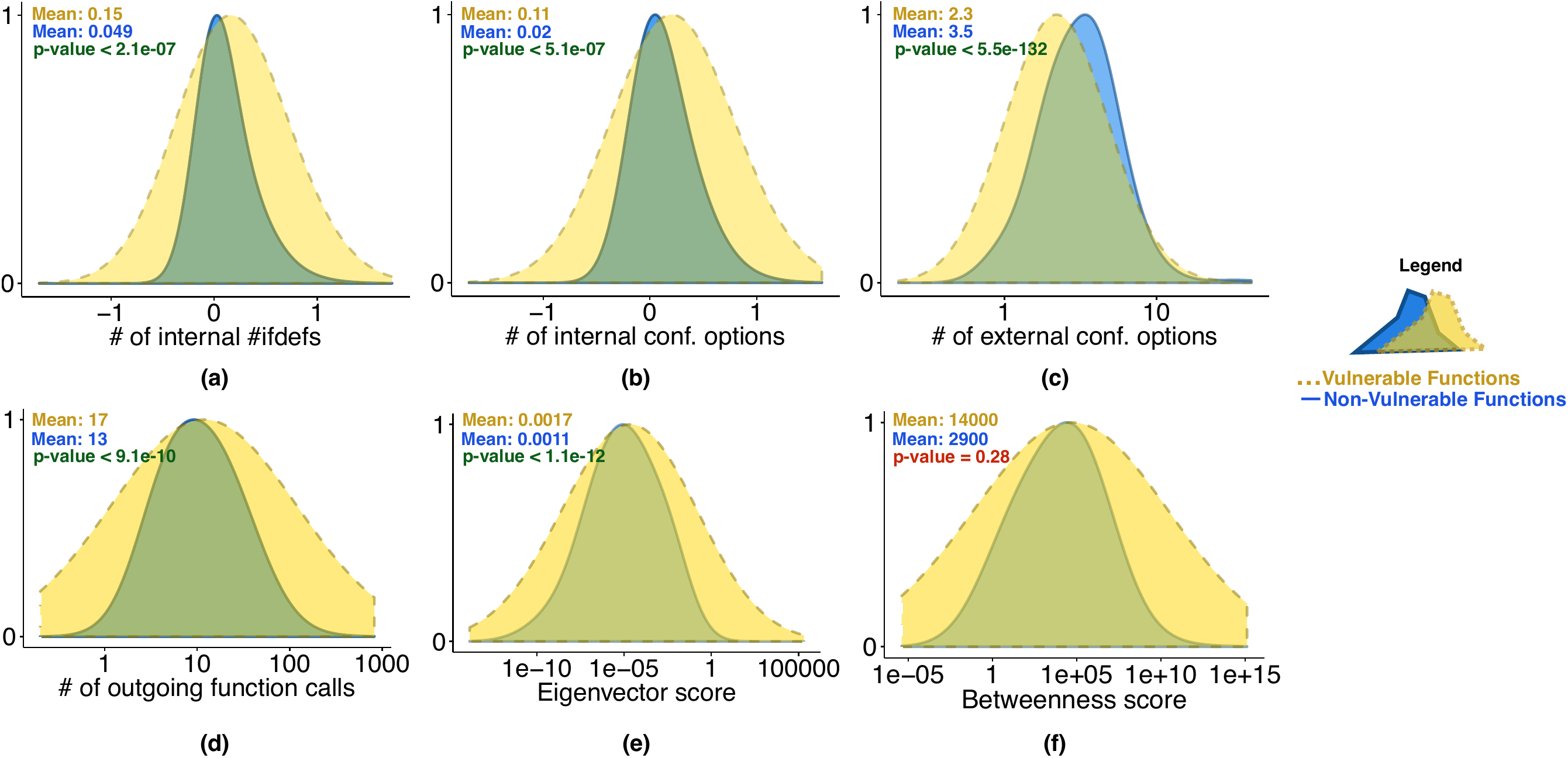}
	\caption{\small\bfseries{Difference of vulnerable and non-vulnerable function samples along configuration complexity (first row) and 
	structural complexity metrics (second row) in $ log_{10} $ scale.}}
	\label{fig:metrics_density}
\end{figure*}

For each metric, we performed a Welch two sample $t$-test between vulnerable and non-vulnerable function samples. We found significant 
differences in distributions and means between vulnerable functions and non-vulnerable functions for many of the selected metrics 
(Figure~\ref{fig:metrics_density}), where vulnerable functions are consistently more complex than non-vulnerable ones 
\footnote{\small{Negative values on the x-axis are a consequence of smoothing the distribution curves for visualization purposes and should 
not be interpreted as valid metrics values.}}.
Also, we report both effect size (difference between means) and statistical significance for each metric, as well as an analysis of 
the validity of our $t$-test statistics (see Appendix. \ref{sec:ttest_validity}). 

In addition to $t$-tests, we applied a confounding effect analysis to check whether our metrics are relevant to characterize complexity, by 
comparing them against existing ones such as size metrics (see Appendix. \ref{sec:confouding_effects}).

\section{Simple Configuration-\\Complexity Metrics}
\label{sec:configuration_comp_analysis}

We define configuration complexity as the complexity induced by the presence of \#ifdefs in source code, and we select a number of metrics 
to quantify it. Quantifying configuration complexity is a challenging task. Our goal is to measure the effect that \#ifdefs and 
configuration options have on developers when they have to understand or change a piece of configurable code as well as how these options 
influence them to make mistakes. To avoid fishing for results, we carefully designed a set of simple metrics that characterize our key 
intuitions of configuration complexity. Each of the following subsections presents an alternative metric to capture configuration 
complexity and its intuition, discusses the results of its null-hypothesis test, and reports an analysis of potential confounding 
effects associated with it.

\begin{figure}[t]
\begin{lstlisting}[language=C, xleftmargin=20pt, frame=bt, basicstyle={\scriptsize\tt}, numbers=left, keywordstyle=\bfseries, 
numberstyle=\scriptsize\tt,  moredelim={[is][\scriptsize]{@}{@}}, deletekeywords={void, int, if}, captionpos=b, belowskip=1pt ]
@#ifdef A@
int foo(int v) {
  int l = read_public_value();
	
  @#if defined(A) || defined(B)
  l = read_private_value();
  @#endif
  
  if (...) {
    v = l + CONST_VALUE; 
  }
  
  @#ifdef C
  assertEquals(v, l + CONST_VALUE)
  @#endif
  
  return v;
}
@#endif
\end{lstlisting}
\caption{\small\bfseries{Example of simple C code with preprocessor directives (\#ifdefs).}}
\label{variational-code}
\end{figure}

\subsection{(Number of) Internal \#ifdefs}
Our first approximation of configuration complexity is to simply count the number of \#ifdefs that appear inside a function. Similar to 
\emph{if} statements, \#ifdef blocks can appear in many different forms in the code, that is, they can be nested, in sequence, or both. 
Our metric captures how many blocks of feature code, regardless of what configuration options are being used, a developer needs to reason 
about while trying to understand or modify a piece of variable code. While it ignores specifics about the variability inside a 
function, it captures the complexity generated by branches of variable code.

For the example, in Figure \ref{variational-code}, there are two \#ifdef blocks inside function \emph{foo} (Lines 5--7 and 15--17), so the 
value of the metric is two. In this example, a developer would need to think about two blocks of variable code, one considering two 
configuration options (A and B, Line 5) and another considering just one (C, Line 15). The insertion of an \#ifdef inside a function, 
whether nested or in sequence to existing ones, would increase the configuration complexity of the code. An increase in the number of 
blocks of feature code inside a function would make a function more complex and more likely to be vulnerable.

\subsubsection*{Results}
	
Our analysis reveals that vulnerable functions have, on average, \functionInternalIfdefsEffectSize{} times more \#ifdefs internally 
(\functionInternalIfdefsVulnerableMean{}) than non-vulnerable functions (\functionInternalIfdefsNonVulnerableMean{}); p < 
\functionInternalIfdefsPValue{}; see Figure \ref{fig:metrics_density}(a).
 
\subsubsection*{Confounding Effect Analysis}	

The correlation coefficient between the number of internal \#ifdefs in a function and its size is \CorInternalIfdefsAndFunctionSize{}, 
which suggests a moderate relationship between the two metrics, that is, long methods often tend to have more \#ifdefs internally. 
When analyzing the regression coefficient for the internal \#ifdefs metric before (\RegressionCoefInternalIfdefs{}) and after the 
size metric is added to the regression model (\RegressionCoefInternalIfdefsAndFunctionSize{}), we see small percentual change in 
the regression coefficient (\DiffRegressionCoefInternalIfdefsWithAndWithoutFunctionSize{} percent), which indicates that there is no 
confounding effect between size and number of internal \#ifdefs.

\subsection{(Number of) Internal Configuration\\ Options}

Complementing the previous metric, our second approximation of configuration complexity counts how many distinct configuration options 
are used within a function. The intuition is that the higher the number of features affecting code inside a function, regardless of 
how many \#ifdefs are in the function, the harder the code is to maintain, due to the increased number of configuration options a developer 
has to consider (remember the number of potential configurations grows exponentially with the number of options). In contrast to our 
previous metric, this metric captures configuration complexity by accounting for the amount of variability inside a function.

For the example in Figure \ref{variational-code}, there are three distinct configuration options used in the two \#ifdef blocks (\emph{A}, 
\emph{B} and \emph{C}) inside the function \emph{foo}, so the value of the metric is three. In this example, a developer would need to 
reason about how three features affect the piece of variable code he is trying to understand or modify. 

\subsubsection*{Results}	

Our analysis reveals  that vulnerable functions have on average \functionInternalIfdefsOptionsEffectSize{} times more configuration options 
internally (\functionInternalIfdefsOptionsVulnerableMean{}) than non-vulnerable functions 
(\functionInternalIfdefsOptionsNonVulnerableMean{}), $p<$\functionInternalIfdefsOptionsPValue{}; see Figure 
\ref{fig:metrics_density}(b).

\subsubsection*{Confounding Effect Analysis}	
The correlation coefficient between number of configuration options used internally and function size is 
\CorInternalConfOptionsAndFunctionSize{}, which suggests a weak relationship between the two metrics. When analyzing the 
regression coefficient for the number of internal configuration options metric before (\RegressionCoefInternalIfdefsOptions{}) and after 
the size metric is added to the regression model (\RegressionCoefInternalIfdefsOptionsAndFunctionSize{}), we see a small change in 
the coefficient (\DiffRegressionCoefInternalIfdefsOptionsWithAndWithoutFunctionSize{} percent), which potentially indicates no confounding 
effect between number of internal configuration options and size.
	
\subsection{(Number of) External Configuration\\ Options}

Different from the two previous metrics that consider how \#ifdef blocks and configuration options affect the complexity \emph{inside} a 
function, our third measure for configuration complexity considers the complexity of the presence condition that constrain the entire 
function. That is, this metric captures the chance that a function is included in a configuration in the first place. It 
counts how many distinct configuration options affect the decision whether a function is  included in a product variant; technically,
it counts the number of options inside \#ifdef blocks \emph{around} the function.\footnote{\small
We do not only consider \#ifdef blocks visible within the .c file, but also conditions from the build system and, often nontrivial, 
interactions among macro definitions, header inclusion, and conditional compilation~ \cite{nadi:2015, kastner:2011}. This analysis is more 
expensive and required significant infrastructure and engineering work, but is also much more precise than just scanning a file for \#ifdef 
directives.}

Our intuition is that the higher is the number of features required to activate a function and its corresponding file, the more complex is 
the condition to activate the code and, consequently, the less often the functionality is included in product variants. 
Functions that are only included in few configurations may be deployed less frequently, thus the chance of finding and exploiting a 
vulnerability is lower; but those functions may also receive less attention in the quality-assurance process, for example, fewer people 
might be interested during code review, leading to a higher chance of vulnerabilities in the future.

For the example of Figure \ref{variational-code}, there is only one configuration option constraining function \emph{foo(A)}, so 
the value of the metric is one. If a function was always included in all product variants, the metric value would be zero, 
since configuration has no effect on the presence or absence of the function. 

\subsubsection*{Results}	

Our analysis reveals  that non-vulnerable functions (\featExprSizeNonVulnerableMean{}) are, on average, constrained by 
\featExprSizeEffectSize{} times more configuration options than vulnerable functions (\featExprSizeVulnerableMean{}), 
$p<$\featExprSizePValue{}; see Figure \ref{fig:metrics_density}(c).

\subsubsection*{Confounding Effect Analysis}	
We expect that the number of external configuration options is independent of the size of a function.
The correlation coefficient between the two is \CorExtenalConfOptionsAndFunctionSize{}, indicating no correlation. 
In addition, when analyzing the regression coefficient for the number of external configuration options metric before 
(\RegressionCoefExternalIfdefOptions{}) and after the size metric is added to the regression model 
(\RegressionCoefExternalIfdefOptionsAndFunctionSize{}), we see a small percentual change 
(\DiffRegressionCoefExternalIfdefOptionsWithAndWithoutFunctionSize{} percent), which potentially excludes a confounding 
effect between the two metrics.

\subsection{Summary}
The results for the simple metrics defined in this section show that they capture distinct characteristics of configuration complexity. 
Despite some limitations of the metrics, including being naturally biased towards syntax rather than semantics and the existence of 
potential confounding size effects, they actually measure distinct characteristics of the variable code. 

While we expected vulnerable functions to have more internal \#ifdefs and more configuration options being used inside them, and 
consequently, to be more complex, we did not expect vulnerable functions to be constrained by fewer configuration options. We can speculate 
this happens because fewer configuration options are required to activate the presence of a function in a product variant and, due to 
broader exposure, more vulnerabilities have been found. Of course, we cannot claim anything about a specific configuration option, but 
assuming that all configuration options have the same chance of being enabled, requiring fewer configuration options would increase the 
chance of a function to be included in a product variant. That is, the chances of a function being exploited would increase along with 
frequency that it is included in product variants of the Linux kernel.

\begin{figure}
	\centering
	\includegraphics[width=.9\linewidth]{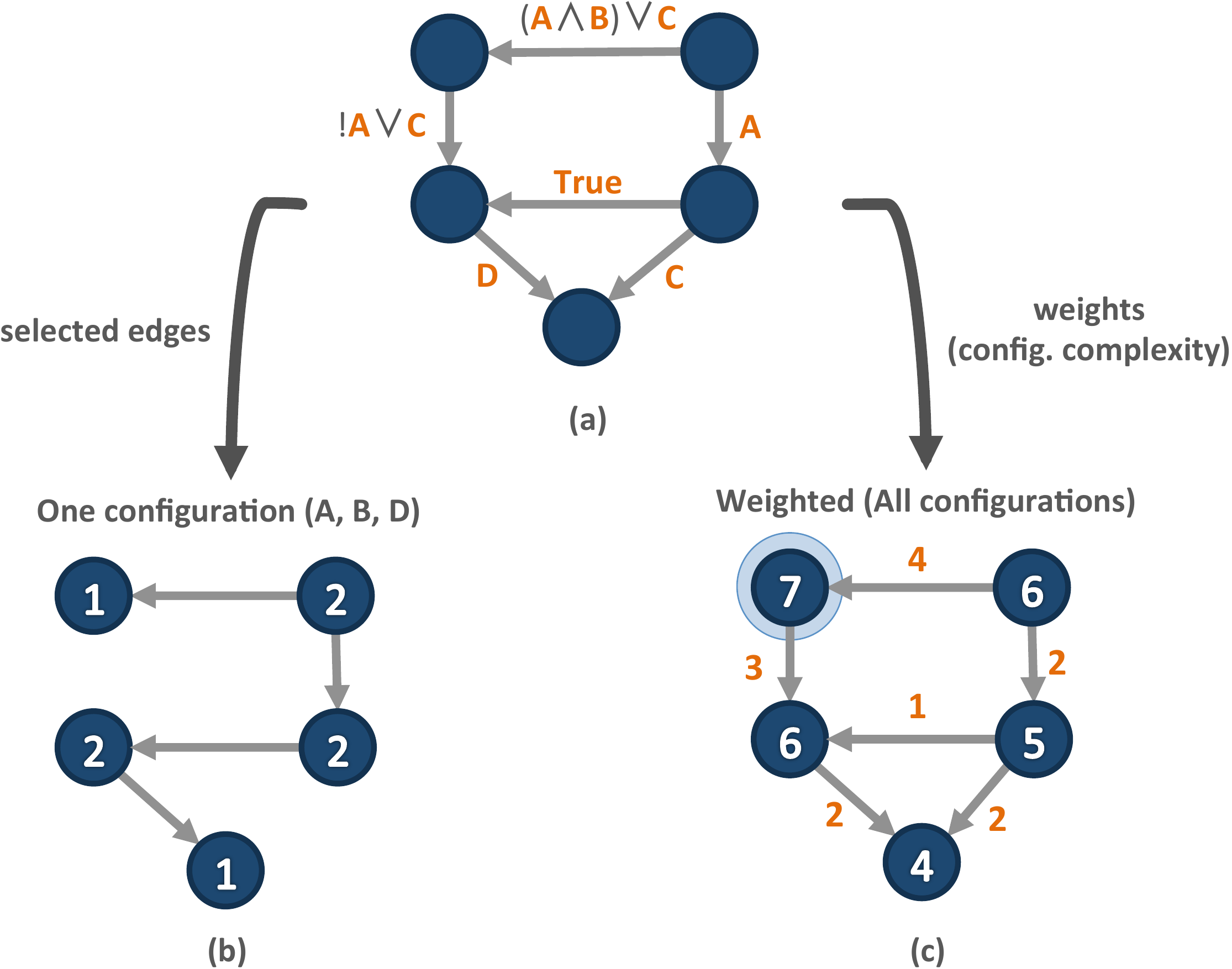}
	\caption{\small\bfseries{Example of a variational call graph with labeled edges representing presence conditions (a); a call graph 
	produced by one specific configuration (when A, B, and D are selected) (b); and the call graph considering all configurations 
	considering the quantification of configuration complexity as weights.}}
	\label{fig:graph_degree}
\end{figure}

\section{Structural Configuration-\\Complexity Metrics}
\label{sec:structural_comp_analysis}

Whereas the previous metrics considered functions in isolation, our three \emph{structural} configuration-complexity metrics characterize
interactions among functions (represented by function calls) and capture how configuration options affect these interactions. All 
structural metrics are based on the variational call graph, introduced in Section \ref{sec:variational_cg}, that compactly describes all 
potential call graphs for all configurations, in which functions (nodes) and function calls (edges) are constrained by presence conditions.

In all three structural metrics that we will present, we transform the presence conditions on edges into weights. Our intuition is that 
calls under very specific conditions are harder to reason about, so we give them a higher weight, roughly similar to a case where we would 
have many different calls between two functions. As weight, we use one plus the number of configuration options that control whether a call 
(edge) is included in a variant. In Figure~\ref{fig:graph_degree}(a,c), we show an example how edge weights are derived from the presence 
conditions in a graph.

Based on previous work \cite{zimmermann:2008, nagappan:2006}, we assume that graph-based metrics are a reliable proxy to measure the 
potential of interaction of nodes in a graph, and consequently, represent their structural complexity. We refine three metrics 
based on standard graph-based metrics \cite{newman:2010} to capture different notions of centrality and, consequently, different notion of 
interactions among functions. The intuition behind these three metrics is that functions that interact, either directly or indirectly, with 
other functions under complicated configuration conditions, are more complex, and consequently, more prone to vulnerable behavior. 
That is, \emph{we create graph-based metrics for configuration complexity based on traditional graph-based metrics by incorporating
weights for configuration decisions and computing them over the entire configuration space}, not just a single product variant.

To separate the configuration aspect from the mechanism of the underlying graph-based metric, we compare each structural 
configuration-complexity metric to a corresponding baseline metric on a single configuration. For example, we compare the 
configuration-weighted eigenvector centrality metric on the call graph for all configurations with an unweighted eigenvector centrality 
metric that we compute on the call graph on a single representative configuration. With this comparison, we can establish whether the 
configuration aspect provides additional information compared to traditional graph-based metrics \cite{newman:2010}; 
Figure~\ref{fig:graph_degree} illustrates that relationship.

As baseline, we use two configurations commonly used for quality-assurance tasks in Linux kernel: the default configuration (\texttt{`make 
defconfig'}) and the maximum configuration (\texttt{`make allyesconfig'}) Especially, the latter is frequently used to increase code 
coverage when testing or analyzing single product variants of the product line \cite{thum:2014}.

\subsection{Degree Centrality}
Our first metric combining structural and configuration-complexity is based on degree centrality~\cite{newman:2010}, which measures the 
immediate importance of a node in the (weighted)  graph by counting how many edges connect that node to other nodes. We consider both 
incoming and outgoing calls and add weights based on the number of external configuration options, as described above. 
We expect that functions with a high configuration-complexity value are called (or calling other functions) often and under complicated 
conditions, and are thus more difficult to understand and more likely to be vulnerable.
Figure \ref{fig:graph_degree} shows how the configuration complexity aspect (represented as weights) changes the result of the metric 
computation compared to a baseline degree centrality metric on an unweighted call graph of a single variant.


\parskip 0pt 
\subsubsection*{Results}	
Our analysis on the complete call graph reveals that vulnerable functions have, on average, a \outDegreeEffectSize{} times more outgoing 
function calls (\outDegreeVulnerableMean{}) than non-vulnerable functions (\outDegreeNonVulnerableMean{}), 
$p<$\outDegreePValue{}; see Figure \ref{fig:metrics_density}(d). That is they call more functions or call them under more complicated 
conditions. The analysis of incoming function calls was not statistically significant.

In comparison, the baseline metric on both the maximum and default configuration does not yield a statistically significant difference 
between vulnerable and non-vulnerable samples, but shows an increased difference of \AllYesConfigOutDegreeEffectSize{} 
and \DefConfigOutDegreeEffectSize{}, respectively for maximum and default configuration. The results show that the addition of 
configuration complexity into the computation of degree amplifies the difference between the two sample means. 


\subsubsection*{Confounding Effect Analysis}	
The correlation coefficients between our metric and the both baseline metrics for the two single configurations (maximum and default) are 
\AllYesConfigCorOutDegreeAndWeightedOutDegree{} and \DefConfigCorOutDegreeAndWeightedOutDegree{} respectively, which suggests a weak 
connection between the two metrics. 

When analyzing the change in odds of the regression coefficients of the weighted out-degree metric before 
(\RegressionCoefWeightedOutDegree{}) and after the out-degree metric from the maximum configuration is added to the regression model 
(\RegressionCoefWeightedOutDegreeAndOutDegree{}), we see a small percentual change of the regression coefficient 
(\DiffRegressionCoefWeightedOutDegreeAndOutDegree{} percent). The weak correlation and confounding analysis results practically excludes a 
confounding effect between the metrics, which shows that considering configuration information improves the 
distinction of vulnerable functions and non-vulnerable ones.

\subsection{Eigenvector Centrality}
Our second structural configuration-complexity metric is based on eigenvector centrality, which is effectively a recursive version of 
the degree centrality, assigning higher values to nodes in neighborhoods of other nodes with high values \cite{newman:2010}.
Again, we compute eigenvector centrality on the weighted call graph of the entire configuration space and compare it against a baseline 
implementation of an unweighted graph of a single configuration. Our intuition is that this metric should be higher for functions with 
complicated conditional call relationships to other functions, especially in neighborhoods where many such complicated call relationships 
exist.
	
\subsubsection*{Results}	
Our analysis on the complete variational call graph reveals that vulnerable functions have, on average, an eigenvector score that is 
\eigenEffectSize{} times (\eigenVulnerableMean{}) higher than non-vulnerable functions (\eigenNonVulnerableMean{}), $p<$\eigenPValue{}.

In comparison, the analysis on both the maximum and default configuration is not statistically significant. The results show that the 
addition of configuration complexity to the computation of eigenvector amplifies the difference between the two sample means and 
highlights the importance of taking configuration complexity into consideration; see Figure \ref{fig:metrics_density}(e).

\subsubsection*{Confounding Effect Analysis}	
The correlation coefficients between our weighted eigenvector score (considering the number of configuration options that compose the 
presence condition on the edges) and the scores for the two single configurations (maximum and default) are 
\AllYesConfigCorEigenAndWeightedEigen{} and \DefConfigCorEigenAndWeightedEigen{}, respectively. Similarly, this suggests a 
moderate correlation between the weighted and unweighted metrics on single configurations, which is expected as they are both 
computed with the same algorithm on similar inputs. 

When analyzing the change in odds of the regression coefficients for the weighted eigenvector metric before 
(\RegressionCoefWeightedEigen{}) and after the eigenvector metric from the maximum configuration  is added to the regression model 
(\RegressionCoefWeightedEigenAndEigen{}), we see, again, a small percentual change of the regression coefficient 
(\DiffRegressionCoefWeightedEigenAndEigen{} percent), which potentially excludes a potential confounding effect between the two 
metrics.

\subsection{Betweenness Centrality}

Our third structural configuration-complexity metric is based on betweenness centrality~\cite{newman:2010}, which captures the notion of 
flow in the graph, an aspect that the two previous metrics do not address. Basically, it computes how many times a node acts as a bridge 
along the shortest path between two other nodes. In the context of variational call graphs, it can be interpreted as the influence 
potential of a function for causing global instability in the call graph. The function with the most strategic location, that is, the 
function that appears in most shortest paths of the call graph, is the most important one; note how this metric approximates the importance 
of a function in the runtime behavior of a program.

To consider configuration complexity, we incorporate the number of external configuration options that constrain the edge, and consequently 
modify the strength of alternative shortest paths (chain of function calls) between two other functions. By considering configuration 
complexity, we intuitively reinforce shortest paths with more complex presence conditions. As baseline, we again compute betweenness 
centrality on the unweighted graph for two single configurations (maximum and default).

\subsubsection*{Results}	
Our analyses on the variational call graph on the maximum, and on the default configuration, are all not statistically significant. For 
this study, the results indicate that betweenness centrality has no sufficient discriminatory power to distinguish vulnerable from 
non-vulnerable functions. We therefore omit analyzing confounding effects; see Figure \ref{fig:metrics_density}(f).

\subsection{Summary}
Overall, we conclude that the configuration complexity aspect of the metrics adds new information to traditional notions of structural 
complexity by amplifying the difference between the metric values for vulnerable and non-vulnerable functions. Our results show that 
combining configuration complexity and structural complexity metrics amplify the observed effect of degree- and 
eigenvector-centrality-based metrics, which signals to be worth paying attention to this combination.

\section{Discussion}
\label{sec:discussion}
We have shown that vulnerable and non-vulnerable functions in the Linux kernel have distinguishable characteristics regarding configuration 
complexity. This result provides a fresh view on the problem of understanding what causes vulnerabilities and whether there are measurable 
correlates that help us in avoiding vulnerabilities. The fact that static variability and preprocessors are widely used in practice has 
been largely ignored in this quest. Our study closes this gap. 

A consequent next step is -- in addition to understanding correlates of vulnerabilities in the presence of static variability -- 
to explore whether we can deduce actionable insights in the form of approved coding guidelines or automatically quantifiable predictors. 
While a thorough treatment is well beyond the scope of this paper,  we will discuss to what degree these insights might be used to predict 
vulnerabilities and guide quality-assurance effort, which additional characteristics might be measured to improve our metrics, and threats 
to validity to our analysis.
		

\subsubsection*{Vulnerability Prediction Challenges}

A persistent modeling challenge is that vulnerable functions are extremely rare in the Linux kernel (\NRVulnerableFunctionsLinuxKernel), 
not giving much information by which to compare them to non-vulnerable functions (\NRNonVulnerableFunctionsLinuxKernel). While we can 
identify differing characteristics (and ensure that they are not caused by the skewness of our data, see 
Appendix.~\ref{sec:ttest_validity}), the difference may not be sufficient to predict at scale. Another issue that we faced is the 
unbalanced nature of the data. That is, in 96 percent of the cases functions do not have \#ifdefs inside their scope. This combined with 
the fact that vulnerabilities are also rare events, make our task of analyzing effect sizes and building predictive models challenging.

We have explored logistic regression and discriminant analysis, but in both cases, the amount of noise and the unbalanced nature of the 
data contributed to a weak prediction model that, in 99 percent of the cases, predicted functions to be non-vulnerable. In that context, 
our metrics make measurable, but effectively tiny improvements to a predictor for vulnerability.

As a meta-result of our study, we arrived at the conclusion that, more than investigating new metrics, we have to develop and apply better 
statistical methods to take the specifics of the data we have at our disposal in to account, in particular, the skewness and availability 
of data.

As said previously, while ending on a sobering note with regard to predictability, our study nonetheless provides novel insights into the 
distinguishing characteristics of vulnerable functions in the presence of static variability -- a dimension that has been overlooked for 
too long. More investigation is required to establish reliable thresholds for these metrics and to improve them to be used in predictive 
models.

\subsubsection*{Refining Configuration Complexity Metrics}	

While we have shown that even simple metrics, such as counting internal \#ifdefs and configuration options used to constrain feature code, 
expose different characteristics of vulnerable and non-vulnerable functions, we expect that there are additional influence factors
that could capture further aspects of configuration complexity.

For example, we could use analyze the importance of individual configuration options. Potentially, we could incorporate information
about how and where configuration options are documented (for example, where in the hierarchy of a feature model~\cite{kang:1990}), how 
much code is affected by a configuration option, and how many developers have touched code a configuration option. With more information 
on configuration options used in practice (e.g., as in a recent study on configuration challenges~\cite{hubaux:2012}), we could even 
characterize how frequently certain configuration options are included in product variants used (and tested but also exploitable) in 
practice. In addition, with information about developers (e.g., developer/code networks~\cite{joblin:2015}), we could identify which 
options have been developed by groups of experienced developers in a domain familiar to them. 

We believe that there are many characteristics of variability left to explore. For instance, in an exploratory analysis, we found that the 
presence conditions of a major part of vulnerable functions usually contain only few configuration options, and that these options are 
often defined near the top of the feature model. This result may corroborate that functions frequently included in the build process are 
being noticed and more frequently screened for vulnerabilities. We think our study on configuration complexity contributes to making 
variability information more accessible. Ultimately, when able to understand complexity metrics and their associated thresholds, we 
envision the creation of a dashboard that aggregates other quantitative information on variability (as discussed in the literature 
\cite{ziegler:2016, joblin:2015, apel:2013, hubaux:2012})  to better support product line maintainability.

\subsubsection*{Threats to Validity}

We acknowledge that we cannot generalize and claim representativeness from our single case study of the Linux Kernel. Nonetheless, we 
have selected the Linux kernel because it is an important case of a product line and also the one with the largest number of reported 
vulnerabilities. For example, OpenSSL has a much smaller code base and only 139 reported vulnerabilities available for analysis.

Our extraction process can potentially threaten the validity of our conclusions. For instance, when investigating the vulnerability history 
of functions, we rely on the vulnerability database completeness and on CVE reports and commits accuracy, which are both produced by humans 
and are consequently subject to human error. Also, we discard information of multiple appearances of a function in the vulnerability 
history and consider only whether a function was once vulnerable or not. This way, we lose potentially important information on 
code churn, but also simplify the analysis. 

We use third-party software to parse and extract simple size metrics from C code (srcML \cite{collard:2003}), and to calculate 
graph-based metrics (igraph \cite{csardi:igraph:2006}). Issues could arise if, for example, the parser is tricked by unusual and obscure 
use preprocessor directives. From prior studies, we know that these cases are rare, though~\cite{liebig:2010}. 

As discussed, our analysis results suffer from the high skewness of the data. The rareness of the events we are interested in, such as, the 
number of vulnerable functions and the number of \#ifdefs used inside functions, required us to be careful when using statistical 
techniques for data analysis. We addressed this as far as possible with corresponding analyses throughout the paper (e.g., checking the 
validity of the t-test in Appendix.~\ref{sec:ttest_validity}). Finally, our configuration-complexity metrics are only proxies for 
actual configuration complexity. For this reason, we explicitly control for potential confounding effects between our metrics and existing 
size and structural complexity ones \cite{fenton:1998, newman:2010}.


\section{Related Work}
Challenges in developing and maintaining variable code with preprocessors are frequently discussed in the literature 
\cite{medeiros:2015,erwig:2011,lohmann:2006}.Researchers state that developers struggle in understanding source code with variability 
because it is hard to keep track of the data-flow and control-flow dependencies and precisely identify what parts of the code are actually 
going to be compiled into a product variant. Medeiros et al.\ \cite{medeiros:2015} have interviewed developers that use the C preprocessor 
in practice and found that they frequently suffer from preprocessor-related problems and 
bugs \cite{abal:2014}. Despite all known challenges, developers do not see alternative technologies that could satisfactorily replace the C 
preprocessor, which indicates that it will continue being used as a main tool to implement variability. Configuration-related 
issues have also been discussed as a severe security threat to software systems \cite{nhlabatsi:2008}.

Similar to our work, Chowdhury et al.\ \cite{chowdhury:2010} investigated the connection between \emph{code} complexity metrics and the 
occurrence of vulnerabilities. Their results suggest that code complexity metrics can be dependably used as early indicators of 
vulnerabilities in software systems. Our work complements their work by defining \emph{configuration complexity} metrics, which capture 
different aspects of complexity, and also in checking whether these metrics can be used as reliable indicators of vulnerabilities. Neuhaus 
et al.\ \cite{neuhaus:2007} investigated the sources of vulnerabilities in software systems. The authors report that components that share 
similar sets of function calls are likely to be vulnerable. We explore this notion by identifying functions that are called under many 
different configuration options and have more complex interactions. More sophisticated metrics have been proposed as an alternative to 
capture complexity of software systems by using graph-based representations \cite{zimmermann:2008, ma:2005}.

One strategy used by many analyses is to simply ignore all configuration-related constructs in the source code and to analyze the system 
after the code has been preprocessed, that is, without configuration information (e.g., either generating a product variant by maximizing 
the number of features enabled or relying on a default configuration) \cite{thum:2014}. Although useful in some cases, since developers 
can reuse existing tools, this strategy produces incomplete results and do not allow them to reason about the configuration options and 
their effects on the system in a systematic fashion. To address this limitation, many researchers recently investigated family-based (or 
variability-aware) analysis across entire configuration spaces \cite{thum:2014}; our mechanism to build variational call graphs is an 
instance of that line of research.


\section{Conclusion}
Preprocessors directives (\#ifdefs) have a bad reputation when maintainability and comprehension are first priorities for product-line 
maintainers. We investigated the influence of configuration complexity on the occurrence of vulnerabilities; our results suggest, among 
others, that vulnerable functions have, on average, three times more internal \#ifdefs than non-vulnerable ones. In addition, vulnerable 
functions are constrained by fewer configuration options, which suggests that developers are inclined to notice functions that are 
frequently compiled in product variants. Our goal is to raise the awareness of developers to handle code variability more systematically, 
since it is an important, but often ignored, aspect of product-line engineering.

\section{Acknowledgments}
This work has been supported by the National Security Agency ``CMU Science of Security Lablet: Composability and 
Usability'' (Award n.: H9823014C0140) and by the the German Research Foundation (AP 206/4 and AP 206/6). G. Ferreira is supported by a 
doctoral research grant from CAPES, Brazil. Grant n. 13713/13-2.


\bibliographystyle{abbrv}
\vspace{-.2em}
{\small{\bibliography{SPLC2016}}}
\balancecolumns

\appendix
\section{T-test Validity}
\label{sec:ttest_validity}
Given the extreme skewness of the data and how few vulnerable functions there are compared to non-vulnerable functions, we were concerned 
about the validity of the $t$-test. We used a bootstrap of the $t$-statistics~\cite{efron:bootstrap:1993} to generate a null distribution 
(taking samples of \NRVulnerableFunctionsLinuxKernel{} from the non-vulnerable functions and performing a $t$-test between the sample and 
the whole, that is, generating a distribution over test statistics for tests where, by construction, the null is true). We found that for 
size, the bootstrap distribution was centered slightly right of zero, but a log transformation to stabilize the variance gave a bootstrap 
distribution that agreed with the theoretical $t$-distribution. For the number of \#ifdefs and configuration options used internally in a 
function (see Section \ref{sec:configuration_comp_analysis}), the bootstrap distribution departed even further from a $t$-distribution, 
and log transformations (with add-one smoothing) helped, but the mode was still right of zero and there was excessive mass at large 
positive 
values. 

Comparing the computed $t$-values against the quantiles of the bootstrap distributions, we found that the difference in means along 
size both before and after a log transformation was significant at the .001 level. The results for the number of \#ifdefs and 
configuration options used internally in a function were significant only at the .01 level in linear scale, but at the .001 level in log 
scale. 

However, we note that, even if the difference in means or log means is statistically significant, it may not be substantively useful, for 
example for building a classifier that uses the difference in means to try and discriminate the two classes, if the difference is not 
strong enough to overcome the massive imbalance in the data (as discussed in Section \ref{sec:discussion}).

\section{Controlling for Confounding Effects}
\label{sec:confouding_effects}
The goal behind controlling for confounding effects is to check whether our metrics are relevant for analysis, that is, to verify if 
they are not redundant in comparison with existing metrics, such as size metrics. For that, we combine two kinds of analysis. 

First, we check the correlation between our metrics (simple and structural configuration complexity) and the baseline metrics (size and 
structural complexity on a single configuration, respectively). 

Next, we replicate the method described by El Eman et al. \cite{elEmam:2001} and use logistic regression models to measure the effect of 
our metrics in the characterization of vulnerabilities when compared to the baseline metrics. This involves computing two logistic 
regression models: one univariate, for the metric of interest, and another multivariate, combining the 
metric of interest and a potentially confounding metric. In both cases, the response variable is vulnerability proneness. If the 
regression coefficient for the studied metric changes substantially when the potentially confounding metric added to the model, it has an 
confounding effect on the metric of interest. Specifically, El Eman exponentiate the coefficient of the metric of interest times the 
standard deviation of the metric of interest, which gives the odds ratio of a one standard deviation increase in the metric of interest 
(instead of a one-unit increase). For example, with a metric, without controlling for size the estimated odds ratio may be 1.15 (15 percent 
increase in odds over even odds), and after controlling for size it is 1.07 (7 percent increase in odds over even odds), they calculate 
this as a 6.96 percent increase. 

However, the magnitude does not give a significance test or tells by how much one variable is confounded by another. Hence, we 
employ a standard statistical test, the analysis of deviance (the logistic regression equivalent to partial $F$-tests) to see if the 
addition of the metric of interest to size is significant. We report the test statistics of the reference distribution for analysis of 
deviance, a chi-squared test, along with $p$-values. An equivalent option would be to make a model with only the metric of interest, check 
its significance, and see if that significance changes with the addition of size.

\end{document}

%% file: SPLC2016_complete_generated.tex
\newcommand\NRFunctionsLinuxKernel{233,903}
\newcommand\NRVulnerableFunctionsLinuxKernel{1,170}
\newcommand\NRNonVulnerableFunctionsLinuxKernel{232,733}
\newcommand\sizeVulnerableMean{55}
\newcommand\sizeNonVulnerableMean{28}
\newcommand\sizeEffectSize{1.9}
\newcommand\sizefIntLow{22}
\newcommand\sizeConfIntHigh{31}
\newcommand\sizePValue{2.3e\textsuperscript{-33}}
\newcommand\inDegreeVulnerableMean{15}
\newcommand\inDegreeNonVulnerableMean{16}
\newcommand\inDegreeEffectSize{0.92}
\newcommand\inDegreeConfIntLow{-14}
\newcommand\inDegreeConfIntHigh{11}
\newcommand\inDegreePValue{0.85}
\newcommand\outDegreeVulnerableMean{17}
\newcommand\outDegreeNonVulnerableMean{13}
\newcommand\outDegreeEffectSize{1.3}
\newcommand\outDegreeConfIntLow{2.7}
\newcommand\outDegreeConfIntHigh{5.2}
\newcommand\outDegreePValue{9.1e\textsuperscript{-10}}
\newcommand\AllYesConfigOutDegreeEffectSize{0.91}
\newcommand\DefConfigOutDegreeEffectSize{0.85}
\newcommand\betweenVulnerableMean{14,000}
\newcommand\betweenNonVulnerableMean{2,900}
\newcommand\betweenEffectSize{5}
\newcommand\betweenConfIntLow{-9,400}
\newcommand\betweenConfIntHigh{32,000}
\newcommand\betweenPValue{0.28}
\newcommand\eigenVulnerableMean{0.0017}
\newcommand\eigenNonVulnerableMean{0.0011}
\newcommand\eigenEffectSize{1.5}
\newcommand\eigenConfIntLow{0.00039}
\newcommand\eigenConfIntHigh{0.00069}
\newcommand\eigenPValue{1.1e\textsuperscript{-12}}
\newcommand\featExprSizeVulnerableMean{2.3}
\newcommand\featExprSizeNonVulnerableMean{3.5}
\newcommand\featExprSizeEffectSize{1.5}
\newcommand\featExprSizeConfIntLow{-1.3}
\newcommand\featExprSizeConfIntHigh{-1.1}
\newcommand\featExprSizePValue{5.5e\textsuperscript{-132}}
\newcommand\functionSizeVulnerableMean{55}
\newcommand\functionSizeNonVulnerableMean{28}
\newcommand\functionSizeEffectSize{1.9}
\newcommand\functionSizeConfIntLow{22}
\newcommand\functionSizeConfIntHigh{31}
\newcommand\functionSizePValue{2.3e\textsuperscript{-33}}
\newcommand\functionInternalIfdefsVulnerableMean{0.15}
\newcommand\functionInternalIfdefsNonVulnerableMean{0.049}
\newcommand\functionInternalIfdefsEffectSize{3.04}
\newcommand\functionInternalIfdefsConfIntLow{0.063}
\newcommand\functionInternalIfdefsConfIntHigh{0.14}
\newcommand\functionInternalIfdefsPValue{2.1e\textsuperscript{-07}}
\newcommand\functionInternalIfdefsOptionsVulnerableMean{0.11}
\newcommand\functionInternalIfdefsOptionsNonVulnerableMean{0.026}
\newcommand\functionInternalIfdefsOptionsEffectSize{4.2}
\newcommand\functionInternalIfdefsOptionsConfIntLow{0.052}
\newcommand\functionInternalIfdefsOptionsConfIntHigh{0.12}
\newcommand\functionInternalIfdefsOptionsPValue{5.1e\textsuperscript{-07}}
\newcommand\CorExtenalConfOptionsAndFunctionSize{0.02}
\newcommand\CorInternalIfdefsAndFunctionSize{0.31}
\newcommand\CorInternalConfOptionsAndFunctionSize{0.17}
\newcommand\CorInternalIfdefsAndInternalConfOptions{0.63}
\newcommand\CorInDegreeAndWeightedInDegree{0.95}
\newcommand\CorOutDegreeAndWeightedOutDegree{0.76}
\newcommand\CorBetweenAndWeightedBetween{0.96}
\newcommand\CorEigenAndWeightedEigen{0.77}
\newcommand\AllYesConfigCorOutDegreeAndWeightedOutDegree{0.0032}
\newcommand\DefConfigCorOutDegreeAndWeightedOutDegree{0.022}
\newcommand\AllYesConfigCorBetweenAndWeightedBetween{0.47}
\newcommand\DefConfigCorBetweenAndWeightedBetween{0.37}
\newcommand\AllYesConfigCorEigenAndWeightedEigen{0.16}
\newcommand\DefConfigCorEigenAndWeightedEigen{0.31}
\newcommand\RegressionCoefInternalIfdefs{7.6e\textsuperscript{-09}}
\newcommand\RegressionCoefInternalIfdefsAndFunctionSize{1.5e\textsuperscript{-10}}
\newcommand\DiffRegressionCoefInternalIfdefsWithAndWithoutFunctionSize{3e\textsuperscript{-07}}
\newcommand\RegressionCoefInternalIfdefsOptions{-8.5e\textsuperscript{-08}}
\newcommand\RegressionCoefInternalIfdefsOptionsAndFunctionSize{-3.5e\textsuperscript{-07}}
\newcommand\DiffRegressionCoefInternalIfdefsOptionsWithAndWithoutFunctionSize{7.6e\textsuperscript{-06}}
\newcommand\RegressionCoefExternalIfdefOptions{-3.2e\textsuperscript{-09}}
\newcommand\RegressionCoefExternalIfdefOptionsAndFunctionSize{-2e\textsuperscript{-09}}
\newcommand\DiffRegressionCoefExternalIfdefOptionsWithAndWithoutFunctionSize{3.7e\textsuperscript{-07}}
\newcommand\RegressionCoefWeightedOutDegree{1.8e\textsuperscript{-10}}
\newcommand\RegressionCoefWeightedOutDegreeAndOutDegree{-6e\textsuperscript{-11}}
\newcommand\DiffRegressionCoefWeightedOutDegreeAndOutDegree{5.3e\textsuperscript{-07}}
\newcommand\RegressionCoefWeightedBetween{2.8e\textsuperscript{-15}}
\newcommand\RegressionCoefWeightedBetweenAndBetween{-1e\textsuperscript{-14}}
\newcommand\DiffRegressionCoefWeightedBetweenAndBetween{2.9e\textsuperscript{-11}}
\newcommand\RegressionCoefWeightedEigen{-6.4e\textsuperscript{-13}}
\newcommand\RegressionCoefWeightedEigenAndEigen{-4.9e\textsuperscript{-14}}
\newcommand\DiffRegressionCoefWeightedEigenAndEigen{1.3e\textsuperscript{-09}}

%% file: SPLC2016.bbl
\begin{thebibliography}{10}
\vspace*{.9em}
\bibitem{abal:2014}
I.~Abal, C.~Brabrand, and A.~Wasowski.
\newblock {42 Variability Bugs in the Linux Kernel: A Qualitative Analysis}.
\newblock In {\em Proc.\ Int'l Conf.\ Automated Software Engineering (ASE)},
  pages 421--432. ACM Press, 2014.

\bibitem{apel:2013}
S.~Apel, S.~Kolesnikov, N.~Siegmund, C.~K\"{a}stner, and B.~Garvin.
\newblock {Exploring Feature Interactions in the Wild: The New
  Feature-interaction Challenge}.
\newblock In {\em Proc.\ Int'l Workshop on Feature-Oriented Software
  Development (FOSD)}, pages 1--8. ACM Press, 2013.

\bibitem{chowdhury:2010}
I.~Chowdhury and M.~Zulkernine.
\newblock {Can Complexity, Coupling, and Cohesion Metrics Be Used As Early
  Indicators of Vulnerabilities?}
\newblock In {\em Proc.\ Symp.\ Applied Computing (SAC)}, pages 1963--1969. ACM
  Press, 2010.

\bibitem{collard:2003}
M.~Collard, H.~Kagdi, and J.~Maletic.
\newblock {An XML-based lightweight C++ fact extractor}.
\newblock In {\em Proc.\ Int'l Workshop on Program Comprehension (IWPC)}, pages
  134--143, 2003.

\bibitem{csardi:igraph:2006}
G.~Csardi and T.~Nepusz.
\newblock {The igraph Software Package for Complex Network Research}.
\newblock {\em InterJournal}, Complex Systems:1695, 2006.

\bibitem{efron:bootstrap:1993}
B.~Efron and R.~Tibshiran.
\newblock {\em {An Introduction to the Bootstrap}}.
\newblock Chapman \& Hall, 1993.

\bibitem{elEmam:2001}
K.~El~Emam, S.~Benlarbi, N.~Goel, and S.~Rai.
\newblock {The Confounding Effect of Class Size on the Validity of
  Object-oriented Metrics}.
\newblock {\em IEEE Trans.\ Softw.\ Eng. (TSE)}, 27(7):630--650, 2001.

\bibitem{fenton:1998}
N.~Fenton and S.~Pfleeger.
\newblock {\em {Software Metrics: A Rigorous and Practical Approach}}.
\newblock PWS Publishing Co., 2nd edition, 1998.

\bibitem{ferreira:2015}
G.~Ferreira, C.~K{\"{a}}stner, J.~Pfeffer, and S.~Apel.
\newblock {Characterizing Complexity of Highly-configurable Systems with
  Variational Call Graphs}.
\newblock In {\em Proc.\ Symposium and Bootcamp on the Science of Security
  (HotSoS)}, pages 17:1--2, 2015.

\bibitem{garvin:2011}
B.~Garvin and M.~Cohen.
\newblock {Feature Interaction Faults Revisited: An Exploratory Study}.
\newblock In {\em Proc.\ Int'l Symp.\ Software Reliability Engineering
  (ISSRE)}, pages 90--99, 2011.

\bibitem{gazzillo:2012}
P.~Gazzillo and R.~Grimm.
\newblock {SuperC: Parsing All of C by Taming the Preprocessor}.
\newblock In {\em Proc.\ Conf.\ Programming Language Design and Implementation
  (PLDI)}, pages 323--334. ACM Press, 2012.

\bibitem{georgiev:2012}
M.~Georgiev, S.~Iyengar, S.~Jana, R.~Anubhai, D.~Boneh, and V.~Shmatikov.
\newblock {The Most Dangerous Code in the World: Validating SSL Certificates in
  Non-browser Software}.
\newblock In {\em Proc. \ Conf.\ on Computer and Communications Security
  (CCS)}, pages 38--49, 2012.

\bibitem{hubaux:2012}
A.~Hubaux, Y.~Xiong, and C.~Krzysztof.
\newblock {A User Survey of Configuration Challenges in Linux and eCos}.
\newblock In {\em Proc.\ Int'l Workshop on Variability Modelling of
  Software-intensive Systems (VaMoS)}. ACM Press, 2012.

\bibitem{hunsen:2015}
C.~Hunsen, B.~Zhang, J.~Siegmund, C.~K{\"a}stner, O.~Le{\ss}enich, M.~Becker,
  and S.~Apel.
\newblock {{Preprocessor-based Variability in Open-source and Industrial
  Software Systems: An Empirical Study}}.
\newblock {\em Empirical Software Engineering}, 21(2):449--482, 2006.

\bibitem{joblin:2015}
M.~Joblin, W.~Mauerer, S.~Apel, J.~Siegmund, and D.~Riehle.
\newblock {From Developer Networks to Verified Communities: A Fine-grained
  Approach}.
\newblock In {\em Proc.\ Int'l Conf.\ Software Engineering (ICSE)}, pages
  563--573. IEEE Computer Society, 2015.

\bibitem{johnson:2009}
N.~F. Johnson.
\newblock {\em {Simply Complexity: A Clear Guide to Complexity Theory}}.
\newblock Oneworld Publications, 2009.

\bibitem{kang:1990}
K.~Kang, S.~Cohen, J.~Hess, W.~Novak, and A.~Peterson.
\newblock {Feature-Oriented Domain Analysis (FODA) Feasibility Study}.
\newblock Technical Report CMU/SEI-90-TR-021, Software Engineering Institute,
  Carnegie Mellon University, 1990.

\bibitem{kastner:2012}
C.~K{\"a}stner, S.~Apel, T.~Th{\"u}m, and G.~Saake.
\newblock {Type Checking Annotation-Based Product Lines}.
\newblock {\em ACM Trans. Softw. Eng. Methodol. (TOSEM)}, 21(3):14:1--14:39,
  2012.

\bibitem{kastner:2011}
C.~K{\"{a}}stner, P.~G. Giarrusso, T.~Rendel, S.~Erdweg, K.~Ostermann, and
  T.~Berger.
\newblock {Variability-aware Parsing in the Presence of Lexical Macros and
  Conditional Compilation}.
\newblock In {\em Proc.\ Int'l Conf.\ Object-Oriented Programming, Systems,
  Languages and Applications (OOPSLA)}, pages 805--824. ACM Press, 2011.

\bibitem{latoza:2006}
T.~D. LaToza, G.~Venolia, and R.~DeLine.
\newblock {Maintaining Mental Models: A Study of Developer Work Habits}.
\newblock In {\em Proc.\ Int'l Conf.\ Software Engineering (ICSE)}, pages
  492--501. ACM Press, 2006.

\bibitem{erwig:2011}
D.~Le, E.~Walkingshaw, and M.~Erwig.
\newblock {\#ifdef Confirmed Harmful: Promoting Understandable Software
  Variation}.
\newblock In {\em Proc.\ Int'l Symp. Visual Languages and Human-Centric
  Computing (VLHCC)}, pages 143--150, 2011.

\bibitem{liebig:2010}
J.~Liebig, S.~Apel, C.~Lengauer, C.~K\"{a}stner, and M.~Schulze.
\newblock {An Analysis of the Variability in Forty Preprocessor-based Software
  Product Lines}.
\newblock In {\em Proc.\ Int'l Conf.\ Software Engineering (ICSE)}, pages
  105--114. ACM Press, 2010.

\bibitem{liebig:2013}
J.~Liebig, A.~von Rhein, C.~K{\"{a}}stner, S.~Apel, J.~D{\"{o}}rre, and
  C.~Lengauer.
\newblock {Scalable Analysis of Variable Software}.
\newblock In {\em Proc.\ Europ.\ Software Engineering Conf./Foundations of
  Software Engineering (ESEC/FSE)}, pages 81--91, 2013.

\bibitem{lohmann:2006}
D.~Lohmann, F.~Scheler, R.~Tartler, O.~Spinczyk, and
  W.~Schr\"{o}der-Preikschat.
\newblock {A Quantitative Analysis of Aspects in the eCos Kernel}.
\newblock In {\em Proc.\ Europ.\ Conf.\ Computer Systems (EuroSys)}, pages
  191--204. ACM Press, 2006.

\bibitem{ma:2005}
Y.~Ma, K.~He, and D.~Du.
\newblock {A Qualitative Method for Measuring the Structural Complexity of
  Software Systems Based on Complex Networks}.
\newblock In {\em Proc.\ Asia-Pacific Software Engineering Conf.\ (APSEC)},
  pages 257--263. IEEE Computer Society, 2005.

\bibitem{mccabe:1976}
T.~J. McCabe.
\newblock {A Complexity Measure}.
\newblock {\em IEEE Trans.\ Softw.\ Eng. (TSE)}, 2(4):308--320, 1976.

\bibitem{medeiros:2015}
F.~Medeiros, C.~K{\"a}stner, M.~Ribeiro, S.~Nadi, and R.~Gheyi.
\newblock {The Love/Hate Relationship with The C Preprocessor: An Interview
  Study}.
\newblock In {\em Proc.\ Europ.\ Conf.\ Object-Oriented Programming (ECOOP)},
  pages 495--518. Springer-Verlag, 2015.

\bibitem{nadi:2015}
S.~Nadi, T.~Berger, C.~K{\"a}stner, and K.~Czarnecki.
\newblock {Where Do Configuration Constraints Stem From? An Extraction Approach
  and an Empirical Study}.
\newblock {\em IEEE Trans.\ Softw.\ Eng. (TSE)}, 41(8):820--841, 2015.

\bibitem{nagappan:2006}
N.~Nagappan, T.~Ball, and A.~Zeller.
\newblock {Mining Metrics to Predict Component Failures}.
\newblock In {\em Proc.\ Int'l Conf.\ Software Engineering (ICSE)}, pages
  452--461. ACM Press, 2006.

\bibitem{neuhaus:2007}
S.~Neuhaus, T.~Zimmermann, C.~Holler, and A.~Zeller.
\newblock {Predicting Vulnerable Software Components}.
\newblock In {\em Proc. \ Conf.\ on Computer and Communications Security
  (CCS)}, pages 529--540. ACM Press, 2007.

\bibitem{newman:2010}
M.~Newman.
\newblock {\em Networks: An Introduction}.
\newblock Oxford University Press, Inc., 2010.

\bibitem{nhlabatsi:2008}
A.~Nhlabatsi, R.~Laney, and B.~Nuseibeh.
\newblock {Feature Interaction: The Security Threat from within Software
  Systems}.
\newblock {\em Progress in Informatics}, 1(5):75--89, 2008.

\bibitem{pohl:2005:SPL}
K.~Pohl, G.~B\"{o}ckle, and F.~J. van~der Linden.
\newblock {\em {Software Product Line Engineering: Foundations, Principles and
  Techniques}}.
\newblock Springer-Verlag, 2005.

\bibitem{pohl:2006:SPLTesting}
K.~Pohl and A.~Metzger.
\newblock {Software Product Line Testing}.
\newblock {\em Commun.\ ACM}, 49(12):78--81, 2006.

\bibitem{post:2008}
H.~Post and Carsten.
\newblock {Configuration Lifting: Verification Meets Software Configuration}.
\newblock In {\em Proc.\ Int'l Conf.\ Automated Software Engineering (ASE)},
  pages 347--350. IEEE Computer Society, 2008.

\bibitem{schulze:2013}
S.~Schulze, J.~Liebig, J.~Siegmund, and S.~Apel.
\newblock {Does the Discipline of Preprocessor Annotations Matter?: A
  Controlled Experiment}.
\newblock In {\em Proc.\ Int'l Conf.\ Generative Programming and Component
  Engineering (GPCE)}, pages 65--74. ACM Press, 2013.

\bibitem{sincero:2007}
J.~Sincero, H.~Schirmeier, W.~Schr{\"o}der-Preikschat, and O.~Spinczyk.
\newblock {Is the Linux Kernel a Software Product Line?}
\newblock In {\em Proc.\ Int'l Workshop on Open Source Software and Product
  Lines (OSSPL)}, 2007.

\bibitem{tartler:2012}
R.~Tartler, A.~Kurmus, B.~Heinloth, V.~Rothberg, A.~Ruprecht, D.~Dorneanu,
  R.~Kapitza, W.~Schr\"{o}der-Preikschat, and D.~Lohmann.
\newblock {Automatic OS Kernel TCB Reduction by Leveraging Compile-time
  Configurability}.
\newblock In {\em Proc. USENIX Conf.\ on Hot Topics in System Dependability
  (HotDep)}, pages 3--3, 2012.

\bibitem{thum:2014}
T.~Th{\"{u}}m, S.~Apel, C.~K{\"{a}}stner, I.~Schaefer, and G.~Saake.
\newblock {A Classification and Survey of Analysis Strategies for Software
  Product Lines}.
\newblock {\em ACM Computing Surveys (CSUR)}, 47(1):6, 2014.

\bibitem{vonRhein:2016}
A.~von Rhein, T.~Th{\"{u}}m, I.~Schaefer, J.~Liebig, and S.~Apel.
\newblock {Variability Encoding: From Compile-time to Load-time Variability}.
\newblock {\em Journal of Logical and Algebraic Methods in Programming},
  85(1):125--145, 2016.

\bibitem{wagner:2001}
D.~Wagner and D.~Dean.
\newblock {Intrusion Detection via Static Analysis}.
\newblock In {\em Proc. Symposium on Security and Privacy (SP)}, page 156. IEEE
  Computer Society, 2001.

\bibitem{ziegler:2016}
A.~Ziegler, V.~Rothberg, and D.~Lohmann.
\newblock {Analyzing the Impact of Feature Changes in Linux}.
\newblock In {\em Proc.\ Int'l Workshop on Variability Modelling of
  Software-intensive Systems (VaMoS)}, pages 25--32. ACM Press, 2016.

\bibitem{zimmermann:2008}
T.~Zimmermann and N.~Nagappan.
\newblock {Predicting Defects Using Network Analysis on Dependency Graphs}.
\newblock In {\em Proc.\ Int'l Conf.\ Software Engineering (ICSE)}, pages
  531--540. ACM Press, 2008.

\end{thebibliography}
